\documentclass[letter,12pt]{article}
\usepackage{graphicx}
\usepackage{authblk}
\usepackage{polski}
\usepackage[english]{babel}

\title{Evidence for weakly bound electrons in non-irradiated alkane crystals. The electrons as a probe of structural differences in crystals.}
\author[1]{M.~Pietrow\thanks{corresponding author; email: mrk@kft.umcs.lublin.pl}}
\author[2]{M.~Gago\'s}
\author[1]{L.E.~Misiak}
\author[3]{K.~Kornarzy\'nski}
\author[4]{J.~Szurkowski}
\author[4,5]{P.~Rochowski}
\author[4]{M.~Grzegorczyk}
\affil[1]{Institute of Physics, M. Curie-Sk{\l}odowska University, ul. Pl. M. Curie-Sk{\l}odowskiej 1, 20-031 Lublin, Poland}
\affil[2]{Department of Cell Biology, Institute of Biology and Biochemistry, Maria Curie-Sk{\l}odowska University, ul. Akademicka 19, 20-033 Lublin, Poland}
\affil[3]{Department of Biophysics, University of Life Sciences in Lublin, Akademicka 13, 20-950 Lublin, Poland}
\affil[4]{Institute of Experimental Physics, University of Gda\'nsk, Wita Stwosza 57, 80-952 Gda\'nsk, Poland}
\affil[5]{Pomeranian University in S{\l}upsk, Arciszewskiego 22b, 76-200 S{\l}upsk, Poland}
\begin{document}
\maketitle
\section{Abstract}
It is generally assumed that weakly bound (trapped) electrons in organic solids come only from radiolytical (or photochemical) processes like ionization caused by an excited positron entering the sample. This paper presents an evidence for the presence of these electrons in non-irradiated samples of docosane. We argue that these electrons can be located (trapped) either in interlamellar gaps or in spaces made by non-planar conformers. The electrons from the former ones are bound more weakly than those from the latter ones. The origin of Vis absorption for the samples is explained. These spectra can be used as a probe indicating differences in the solid structures of hydrocarbons.
\paragraph{Keywords:} electron traps, crystals of alkanes, molecular crystals, Vis spectroscopy, Positron Annihilation Lifetime Spectroscopy (PALS)
\paragraph{PACS:} 72.90.+y, 33.20.Kf, 36.10.Dr, 61.80.-x, 33.15.Kr
\section{Introduction}
One of the interesting topics related to the processes of energy transfer during radiolysis in solids is existence of the electron trapping process that allows restoring a part of long living products of radiolysis. Quasi-free electrons are produced when the energetic particle enters the sample and loses its energy in favour of ionization of surrounding particles. These electrons can be localized for sufficient time in sites called traps and are accessible for other processes. Because the bond energy in the trap is relatively small (about 1~eV), the electrons can influence chemical reactions as quasi-free radicals. On the other hand, the bond energy is higher than the mean thermal energy at ambient temperature (or lower); therefore, the electrons are stabilized in traps and can be accumulated during the time of radiolytic irradiation.\\
In particular, the trapping process plays an important role during the radiolytic processes induced by the energetic positron ($\beta^{+}$ radiation) penetrating the solid sample. Besides the basic knowledge of the radiolytical processes, the knowledge of the trapping mechanism allows proper interpretation of many results of experiments supported by positron-based techniques. Especially, the processes related to trapped electrons are the subject of investigations in the PALS technique (\emph{Positron Annihilation Lifetime Spectroscopy}) \cite{Hirade10}.\\
The most useful medium for investigating the trapping processes here are simple polymers and alkanes due to their weakly polarized bonds. Here, there are no fractionally charged sites, which could catch the electrons and decrease the process of accumulation thereof in shallower traps.\\
The subject of investigations of traps in alkanes in positron annihilation studies are traps that can collect electrons for a long enough time to facilitate catching them by positrons (the typical positron lifetime in solids are hundreds of picoseconds or nanoseconds in the case of the bound state of the positron called positronium). The electrons can settle in pre-existing or dynamically made traps somewhere in the structure. The lifetime of the electrons in these traps is sufficiently long for accumulation thereof increasingly after each next passage of the positron through the sample. In the meantime, some other processes involving positrons and electrons take place. Namely, when a positron reaches the thermal energy, it is possible for it to capture one of electrons from the surroundings and either annihilate it (\emph{free positron annihilation}) or create the bound state called positronium (Ps). An electron that can react with the positron can be a trapped electron. PALS spectroscopy is a technique which allows analysis of some sample properties \cite{Hirade10} with the use of decomposition of the positron lifetime spectrum into channels related to free positron and ortho- and para-positronium states. The knowledge of the role of the positronium yield related to trapped electrons plays a crucial role in interpretation of PALS experimental results.\\
Besides this fact, the knowledge of storage of electrons in the traps in alkanes is possibly an interesting problem from the point of view of chemical reactions of electrons in biological molecules. Alkanes are the simplest organic chain compounds and the basal structures for such derivatives as alcohols, lipids, and other molecules of the greatest importance in biology.\\
The commonly accepted origin of the electrons for trapping are (photo)chemical and radiolytical processes \cite{Mozumder}. In our previous work, \cite{Pietrow13} we showed that in the case of alkane crystals such electrons seem to exist before radiolysis. This work extends the results presented in that paper. Here, we give further evidence for the presence of these electrons and try to define the electron traps with the use of some experimental techniques. On the other hand, we propose a theory that explains absorption Vis spectra recorded for our samples. This theory is supported by the supposition that the samples contain trapped electrons.
\section{Experimental}
The crystals used in our experiments were made by recrystalization of n--docosane purchased from Sigma-Aldrich. The purity of the stock sample was 99\% (real purity of obtained crystals was considered in \cite{Pietrow13}). We examined three forms with different morphology called by us 'dendrites', 'plates', and 'powder' \cite{Pietrow13}.\\
The optical spectroscopic studies were performed with the use of Cary 300 Bio and Cary 50 spectrophotometers by Varian. Every time, particularly when polarization of incident light was changed, the baseline was measured according to given conditions to correct next measurements.\\
The Raman spectra were measured by a Renishaw inVia Reflex spectrometer.\\
We used a Veeco (USA) NanoScope V AFM microscope and a Nikon AZ100M optical confocal microscope.\\
ESR spectra were recorded with an EPR spectrometer SE/X-2547 (Radiopan, Pozna\'n) equipped with a resonant cavity CX-101TE$_{102}$ and working in the X band ($\sim$9.5~GHz). Measurements were made under the following conditions: microwave power was $\sim$15~mW, modulation frequency 100~kHz with the amplitude 0.6~mT, time constant 0.3~s, scan time 2~min, scan range 50-450~mT. The Varian Instruments
standard weak pitch EPR sample 904450-02, 0,00033~\% pitch in KCl was used as a reference probe to determine unpaired spins in docosane samples.\\
The samples of docosane were placed in a thin-walled tube, precisely made from synthetic quartz (733-SPQ-7 by Wilmad). The synthetic quartz allows the ultraviolet radiation to 200~nm to pass and facilitates measurements of weak ESR signals.\\
Photoacoustic measurements were made with a lab-built, single-beam, photoacoustic spectrometer similar to that described in detail elsewhere (see \cite{Szurkowski} as an example). Modulated measuring light was produced with a mechanical chopper used with a xenon arc lamp and a monochromator. Pressure fluctuations in the closed photoacoustic cell were detected using a microphone.\\
An electric field was applied during the optical spectroscopy measurements by keeping the sample between two conductive indium tin oxide coated glass slides (by Sigma-Aldrich) where a voltage from 0~V to 300~V was supplied. In the case of maximum voltage applied, the field energy of the value 10$^{-5}$~eV per molecule was reached.\\
The constant magnetic field was applied by keeping the sample during the optical spectroscopic measurements in the adapter between two neodymium magnets. The distance between the sample and the magnets was changed to some extent. The field intensity was measured by an SMS-102 magnetometer (by Asonik) in the place of the sample. The maximum value of the field was 12~mT.\\
The procedure of charging the hydrocarbon liquid was as follows. Our initial task was to charge the internal wall of the beaker by the electrons. Due to this, we circumvented its internal and external walls with aluminium foil and charged the external foil with positive charge (a glass rod rubbed with paper). Next, we charged the internal foil with negative charge (electrons) introduced by an ebonite rod rubbed with fur. Because of the attraction created by the positive electric charge from outside, the electrons from the internal foil were more likely to adhered to the glass wall of the beaker. After some cycles of charging, we first removed the internal and then the external foil and some charge remained on internal glass wall. Next, we poured liquid heptadecane contacting it with the charged wall of the beaker. Then, we charged the external foil with the negative charge in order to repeal the electrons from the wall toward the liquid. After shaking the beaker to solve more electrons in the liquid, we poured the charged liquid into the quartz cuvette and put it into a UV-Vis spectrometer. A similar procedure was employed in the case of charging the internal wall by the positive charge. The sample was measured at the temperature 20$^{\circ}$C above the melting point after about one minute after placing the sample in the spectrometer.
\section{Theoretical considerations}
\label{subsec:Theo}
\begin{figure}
\centering
\includegraphics[scale=0.7]{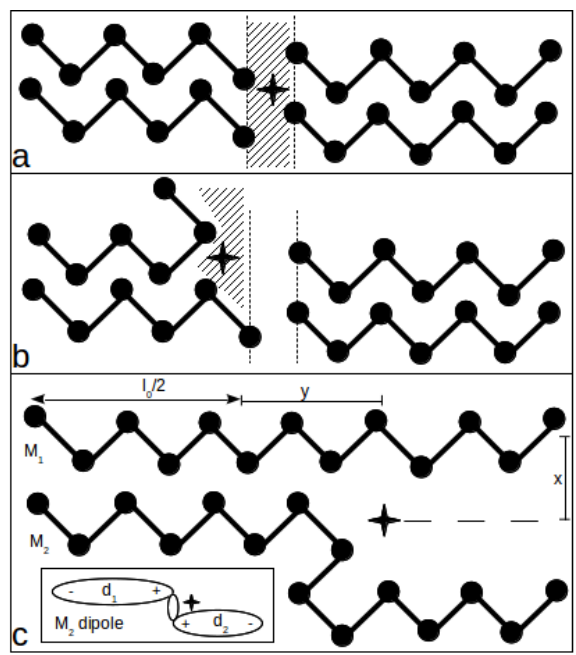}
\caption{Some types of induced electron traps (electron shown as a star): a- the electron in the interlamellar gap (marked region), b- the electron in the cave made by 'end-gauche' conformation of one of the molecules, c- the cave is formed by 'kink' conformation of the molecule. Alphabetic symbols are described in the text. The smaller part of this figure shows division of the distorted molecule into induced dipoles.}
\label{fig:Energia}
\end{figure}

The existence of peaks in the Vis spectra presented here is not obvious but we hope the following theory concerning their origin is reliable.\\
Because alkanes are perfect insulators it is not hard to imagine that they are able to store some electrons from the environment (we have not controlled the charge amount collected on the beaker during the crystallization process). After crystallization, this charge is caged in the crystal structure. Suppose the electron is initially located between molecules and near the interlamellar gap and interacts weakly with the molecules. Let us try to estimate this interaction energy.\\
The polarizability of docosane is ca. $\alpha$=40$\cdot$10$^{-24}$~cm$^3$ \cite{Handbook}. This means that a dipole moment $\mu=\alpha\cdot E$ can be induced on a molecule, where $E$ is the electric vector along the molecule. In our case, $E$ comes from the electrostatic field of the quasi-free electron near one end of the molecule. In the case of an electron in the interlamellar gap, we assume that it is stuck to the positively charged end (the distance $R_w$) and the distance from the other end is equal to the length of the molecule, $l_0$.\\
The dipole moment value can be approximated by taking the electric field as a field from the electron placed $l_0/2$ from the molecule's centre. We have
\begin{equation}
\mu=\alpha\cdot E=\frac{k\ e\ \alpha}{(l_0/2)^2},
\end{equation}
where $e$ is the electron charge, $k=1/(4\pi\epsilon_0)$, and $\epsilon_0$ is the vacuum permittivity.\\
Compare the calculated $\mu$ to the definition of dipole moment $\mu=q\cdot l_0$ and calculate the effective charge separated along the molecule
\begin{equation}
q=\frac{4k\ e \alpha}{l_0^3}.
\end{equation}
Inserting this to the formula for charge-dipole interaction energy yields
\begin{equation}
\frac{k(-e)(-q)}{r_{-}}+\frac{k(-e)(+q)}{r_{+}}=\frac{4k^2e^2\alpha}{l_0^3}(1/l_0-1/R_w).
\end{equation}
Substituting the length of the molecule $l_0$=30~$\AA$ \cite{Wentzel} and $R_w$=1.5~$\AA$, which is an approximate van der Waals radius for carbon and hydrogen atoms, the formula gives -0.054~eV.\\
Thus, the binding energy in a trap of an induced dipole exceeds the mean thermal energy, which is 0.025~eV at room temperature. The electron bond strength is increased by the presence of other dipoles from the nearest neighbourhood. The distance from one chain of the molecule to the other is about 5~$\AA$ (the size of the unit cell for neighbouring C$_{23}$H$_{48}$ was provided in \cite{Wentzel}). It can be assumed that the electron prefers location between chain ends of three molecules (higher value of interaction energy; fig.~\ref{fig:Energia}a) and the interaction energy triples.\\
If the electron settled in the free volume created by such a chain distortion as the kink conformer \cite{Goworek11}, see fig.~\ref{fig:Energia}c, the energy from the dipole generated by all-trans molecule M$_1$ and the energy of interaction with the distorted molecule M$_2$ should be calculated. In the first case, the interaction gives the energy -0.026~eV if $x$=2~$\AA$ and $y$=10~$\AA$ is assumed (the projection of the electric field vector on the molecule axis is small). However, when calculating the interaction with M$_2$, it can be assumed that two dipoles, d$_1$ and d$_2$, interact effectively with the electron. For simplicity, we can assume that the nearest ends are head to head to the electron. The influence of the remaining part of the molecule is negligible. Since the polarizability of the hydrocarbon chain with ten carbon atoms is about two times lower than that for docosane and the distance of the electron to the centre of the dipoles is lower, it can be calculated that the energy of the interaction with one of the dipoles is -0.27~eV for the length of this dipole equal to ten carbon atoms (the kink is located in the centre of the molecule). For non-symmetric kinks the interaction energy rapidly increases and for the length of one of the parts equal to 8$\AA$ the energy is -1.2~eV. Alternatively, for example, two dipoles of the length 10$\AA$ give the energy -1.3~eV.\\
Although the calculations are approximate, they show that the electron binding energy in interlamellar gap is lower than the energy when the electron is in the free volume made by the conformer (mostly for non-symmetric division). The calculated energies are in the agreement with known energies needed to detrap electrons in hydrocarbons \cite{Pietrow}.\\
Accordingly, Vis peaks are expected to appear, and the spectra can be explained by the theory that the electron absorbs energy and migrates between such induced traps with different energies. Most transitions occur between the interlamellar gaps and the volumes made by end-gauche conformers (they are most frequent and the only conformers present at low temperatures~\cite{Maroncelli}; the energy gap between these two traps seems to be adequate to the region of absorbed energy). The transition from trap to trap should be related, in general, to the loss of energy in the structure because of transition work (mainly by lattice vibrations). However, in the case of traps shown in fig.~\ref{fig:Energia}a,b, both free volumes (initial and final position of the electron) are connected with each other and no phonon excitations are expected during the transition.\\
The absorption spectra for the extra electrons that we observed here differ from those observed for electrons released during radiolysis, because in the case of radiolysis recombination with positive ions after release from traps is preferred (lower energetic barrier for transition toward an ion and a recombination with it). In turn, in the case of the extra electrons here, there are no positive ions, and other free volume traps are the preferable sites settled during deexcitation. In this case, energy $W+|\Delta E|$ is absorbed, where $W$ is the trap-to-trap transition work, $\Delta E$- difference between the binding energy inside the initial and a final traps. For the traps in the nearest neighbourhood of each other (no bulk between them), $W$ tends to have the zero value.
\section{Results and Discussion}
\begin{figure}
\centering
\includegraphics[scale=0.4]{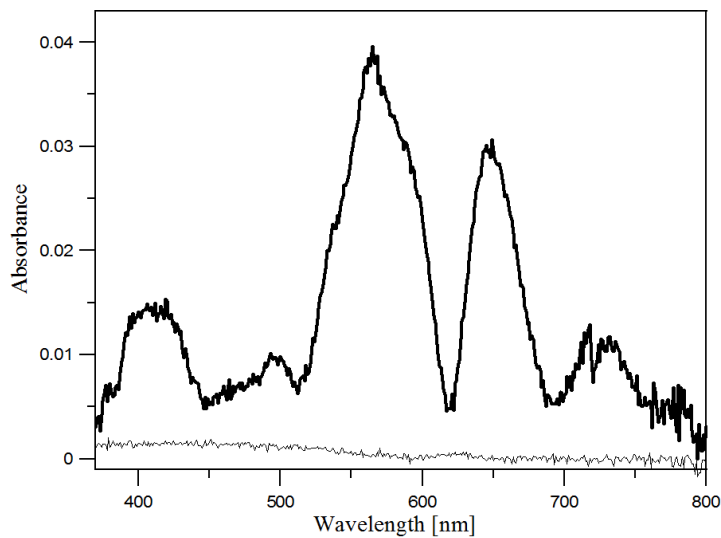}
\caption{Optical absorbance at room temperature for dendrites (bold line) and for a solution of docosane in heptane at the concentration of ca. 30\% (light line). Data in the figure taken from \cite{Pietrow13}.}
\label{fig:UVVis}
\end{figure}

\begin{figure}
\centering
\includegraphics[scale=0.65]{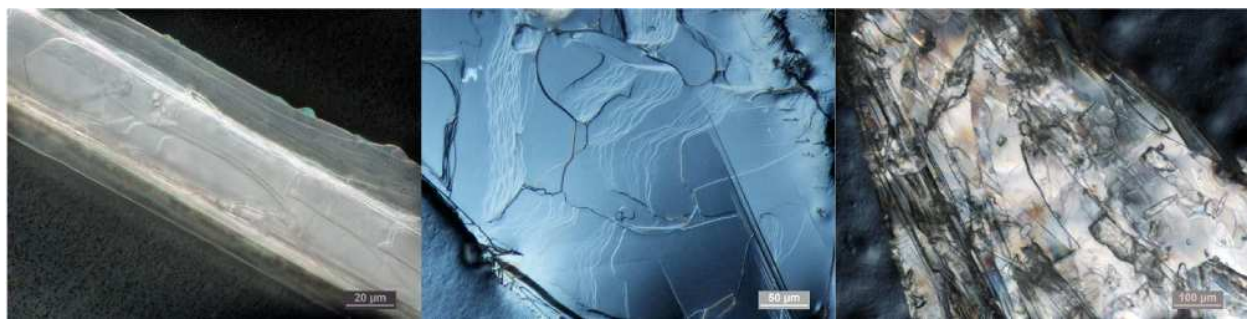}
\caption{Confocal microscope pictures of the samples. From left to right: dendrites, plates, powder.}
\label{fig:OpticalPics}
\end{figure}

\begin{figure}
\centering
\includegraphics[scale=0.64]{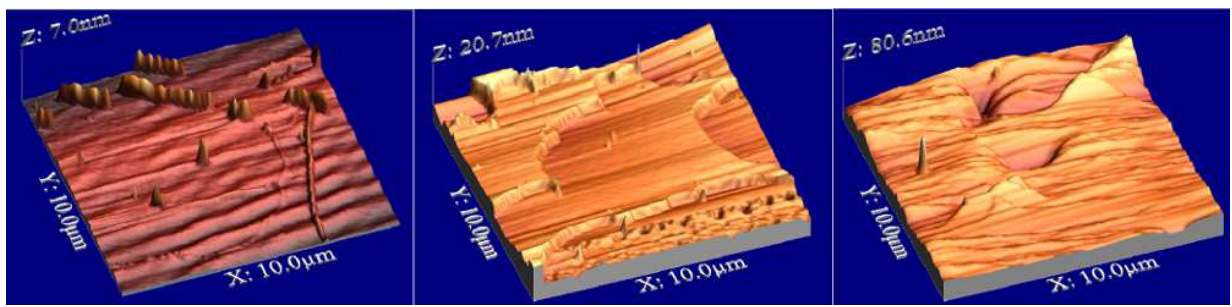}
\caption{AFM pictures of dendrites, plates, powder (from left to right); a 10~$\mu$m square is visible. The scans are visualized by WSxM software \cite{AFMsoft}.}
\label{fig:AFM10um}
\end{figure}

\begin{figure}
\centering
\includegraphics[scale=0.64]{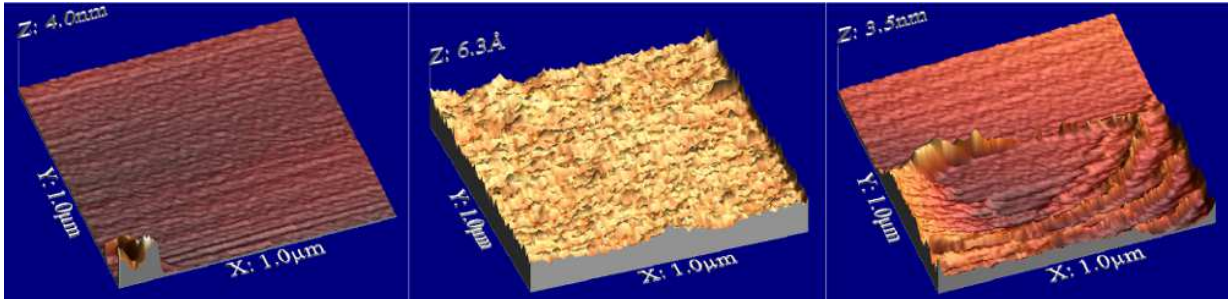}
\caption{AFM pictures of dendrites, plates, powder (from left to right); 1~$\mu$m square is visible. The scans are visualized by WSxM software \cite{AFMsoft}.}
\label{fig:AFM1um}
\end{figure}

\begin{figure}
\centering
\includegraphics[scale=0.65]{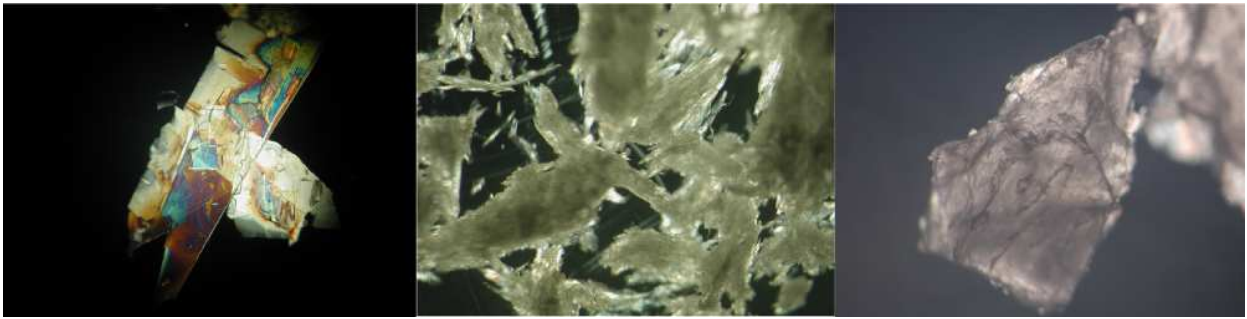}
\caption{The photos of dendrites, plates, powder (from left to right) in polarized light. Magnification 10 times.}
\label{fig:InPolarizedLight}
\end{figure}

\begin{figure}
\centering
\includegraphics[scale=1]{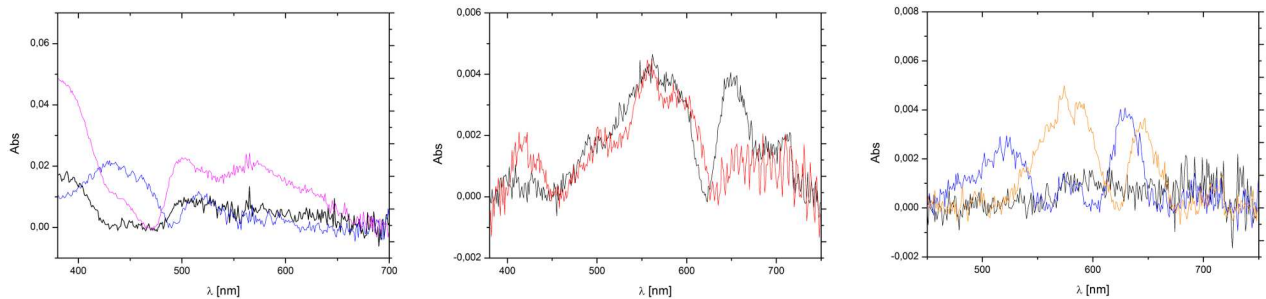}
\caption{From left to right: the VIS spectra of the dendrites in constant electric field (black-~0~V, blue-~0~V after application of 200~V, magenta-~at 300~V), constant magnetic field (black-~without magnetic field, red-~in a field 10~mT), and polarization (black-~polarization angle 0$^{\circ}$, blue-~20$^{\circ}$, orange-~110$^{\circ}$).}
\label{fig:VISchanges}
\end{figure}

\begin{figure}
\centering
\includegraphics[scale=0.5]{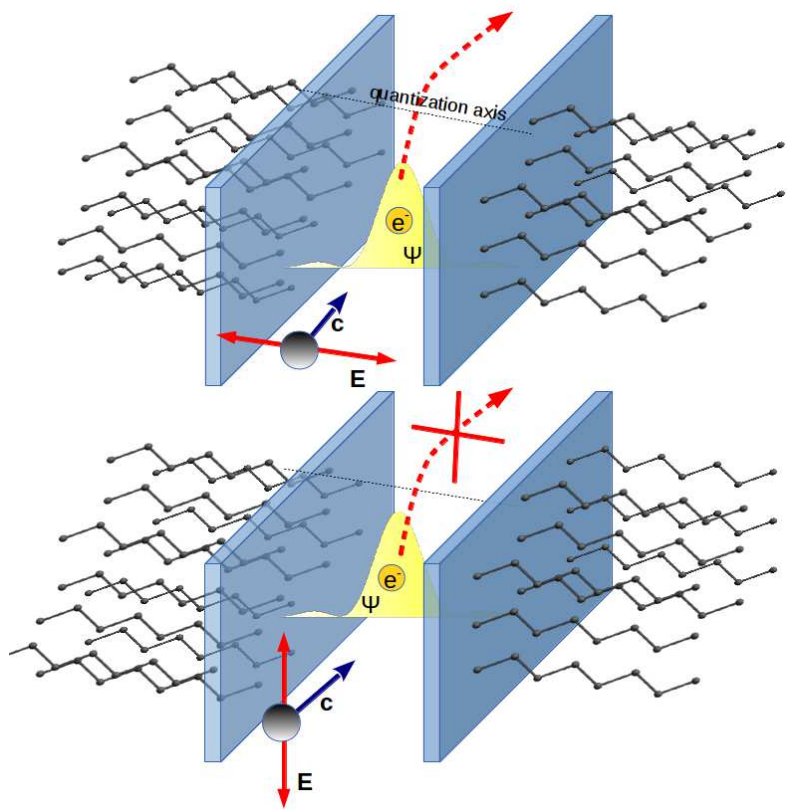}
\caption{The relation between light electric vector $\vec{E}$ orientation (red arrow) and the quantization axis in the electron trap. In the first case, the electron can be excited, whereas in the second one cannot.}
\label{fig:Polar}
\end{figure}

\begin{figure}
\centering
\includegraphics[scale=0.37]{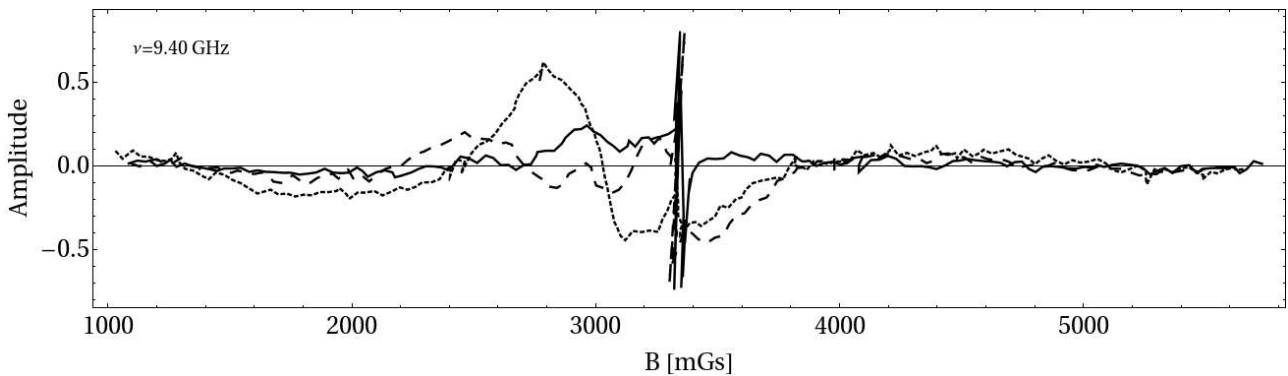}
\caption{ESR spectrum for: dendrites (solid line), plates (dashed line), powder (dotted line). The signal from the DPPH marker (the central line in the figure) exceeds the frame and is cut, which is the cause of 'splitting' the line in the middle into several.}
\label{fig:ESR}
\end{figure}

\begin{figure}
\centering
\includegraphics[scale=0.4]{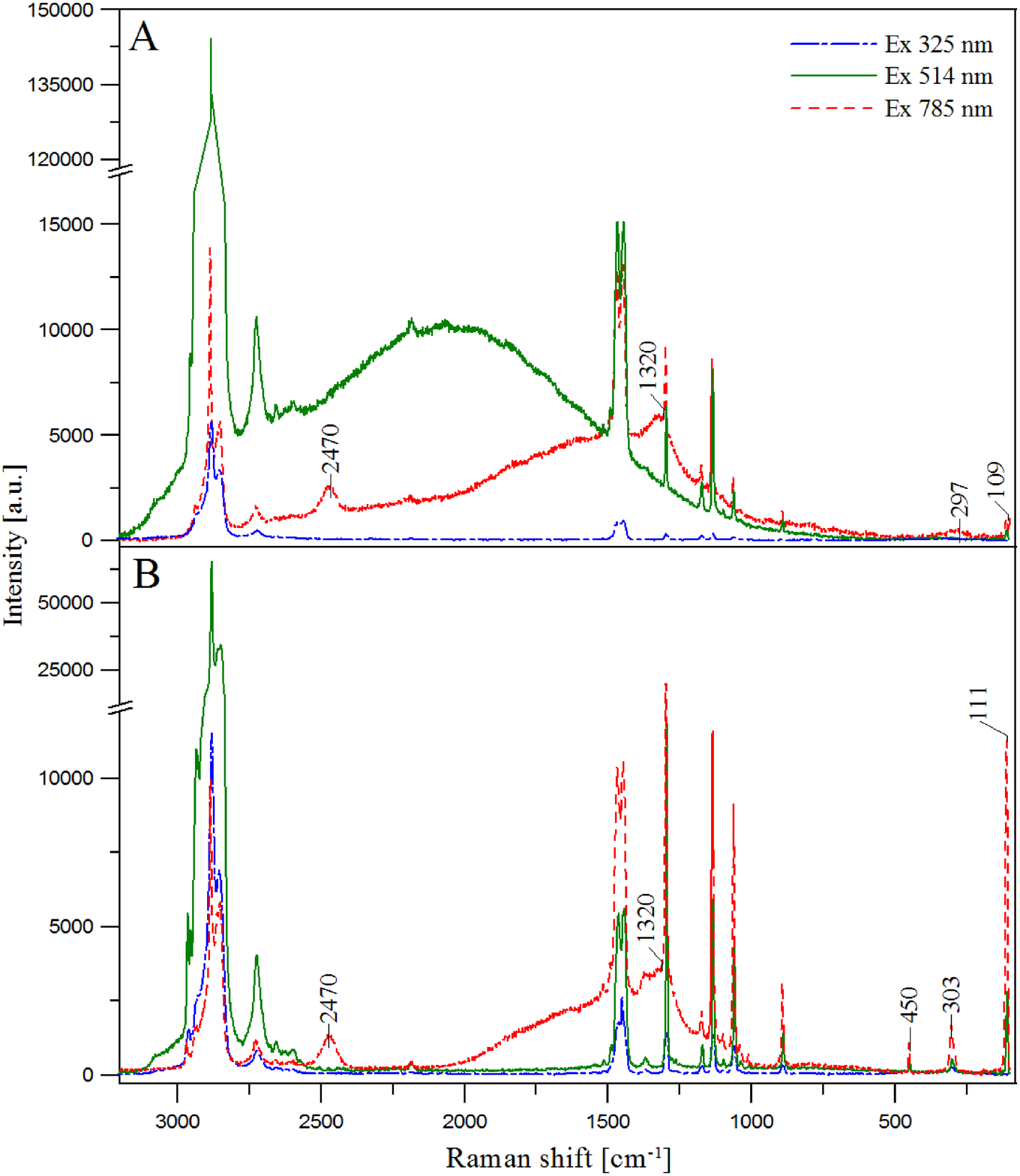}
\caption{Raman spectra for plates (A) and dendrites (B). Red line- 785~nm, green line- 514~nm, blue-~325~nm excitation wavelength.}
\label{fig:Raman}
\end{figure}

\begin{figure}
\centering
\includegraphics[scale=0.5]{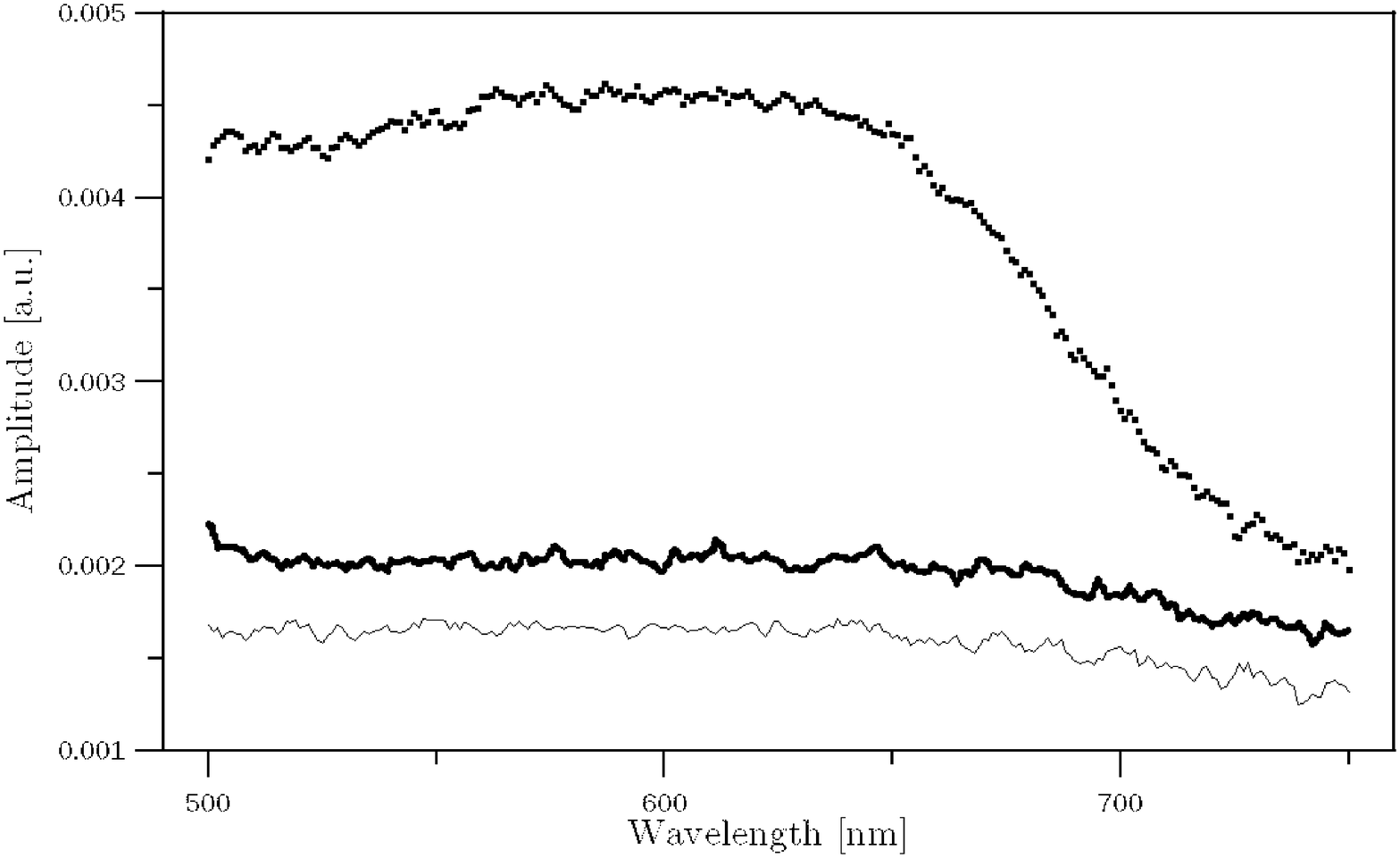}
\caption{Amplitude of the photoacoustic signal against the wavelength of exciting light. The chosen depth (sample excitation region) is set by the modulation frequency 20~Hz. Bold line -- dendrites, thin line -- plates, dots -- powder. Amplitude given as the ordinate is a ratio of the deexcitation energy transformed into heat to the energy of light injected into the sample.}
\label{fig:PAS1}
\end{figure}

\begin{figure}
\centering
\includegraphics[scale=0.47]{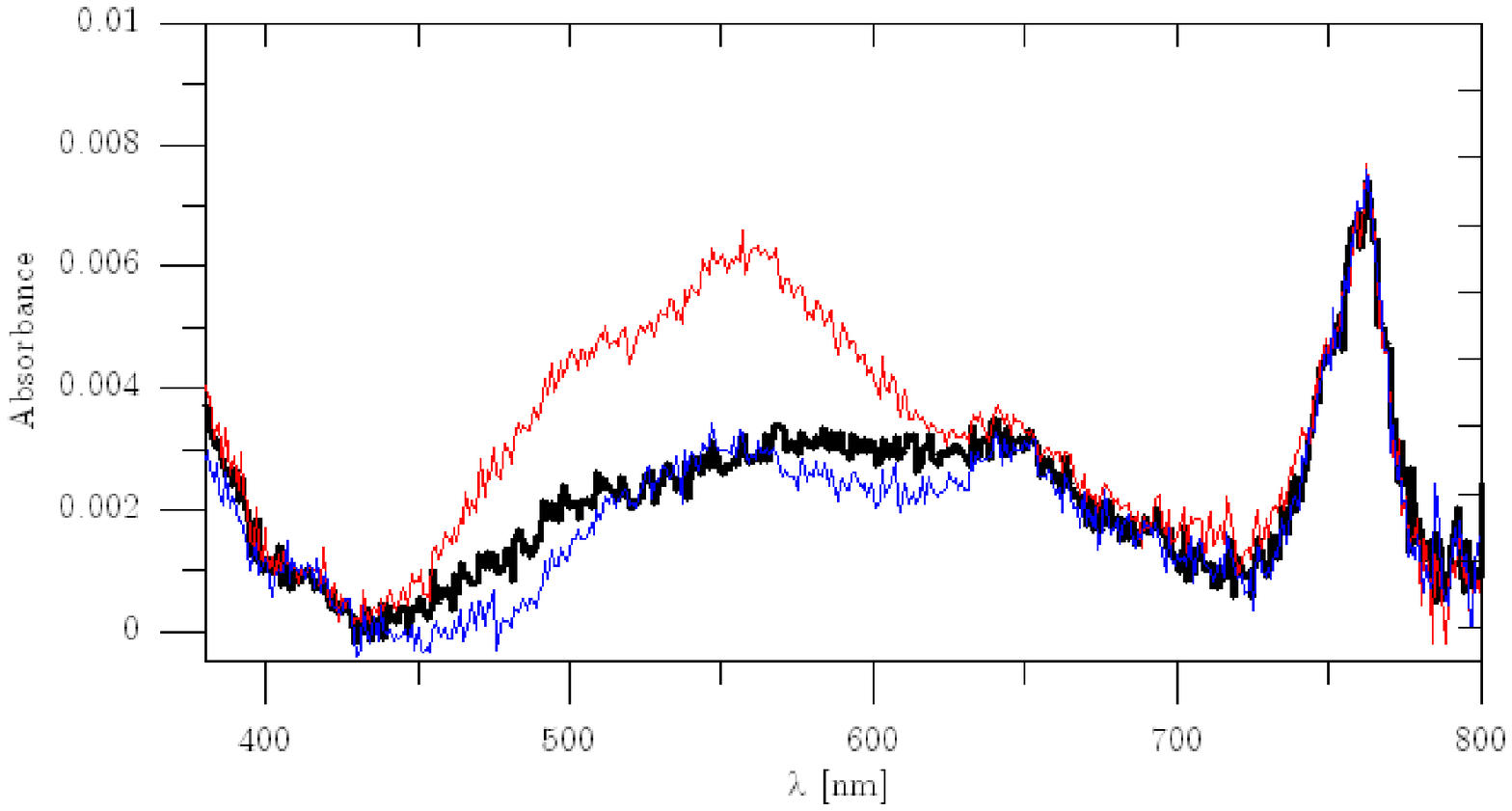}
\includegraphics[scale=0.47]{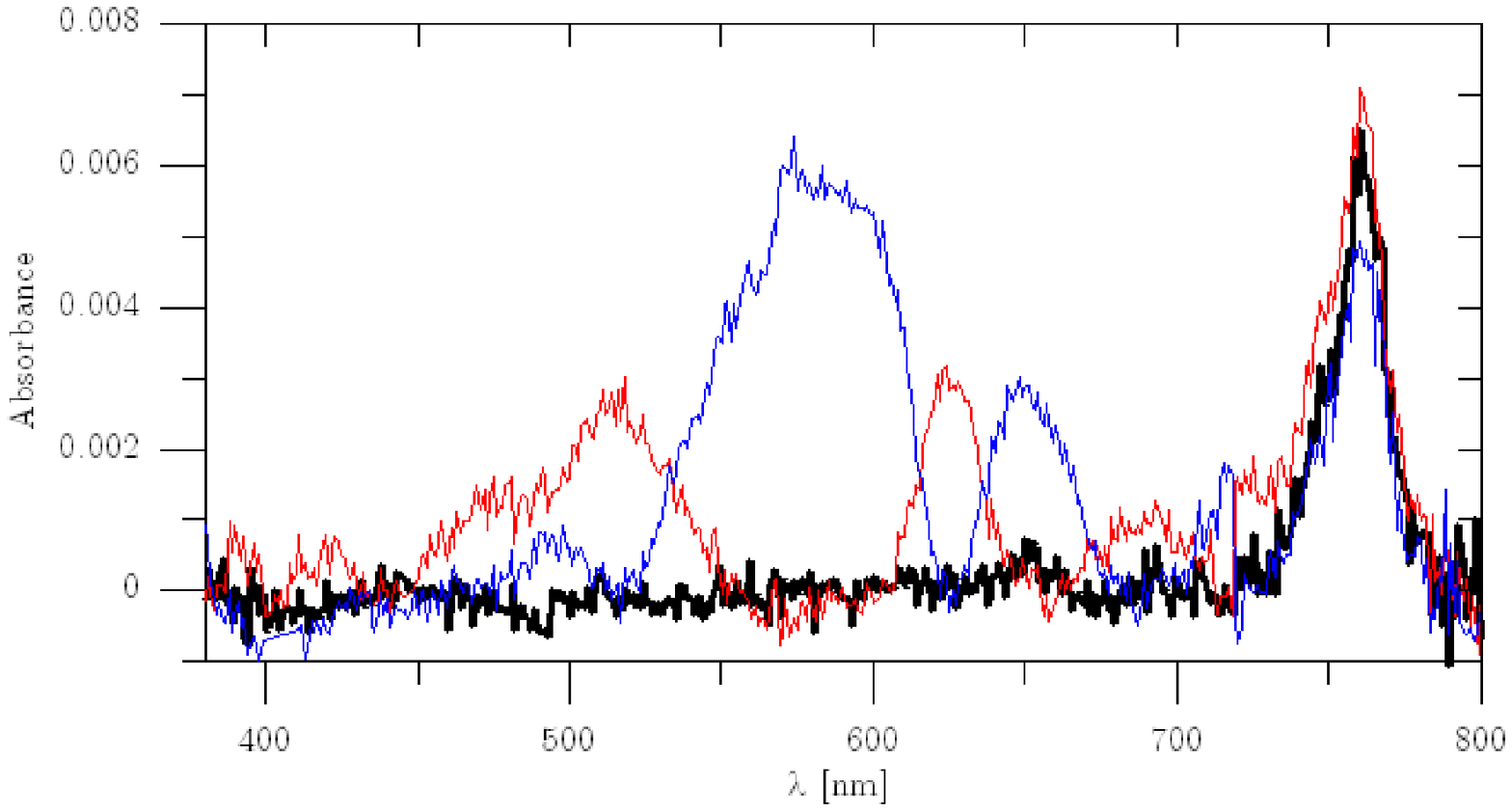}
\caption{Charging the samples: left -- the sample of heptadecane at 50$^{\circ}$C when the negative (red line) and positive charge (blue line) were introduced, compared to the case of the non-charged sample (bold line). Right  -- docosane at 70$^{\circ}$C; non-charged case -- bold line, charged sample -- red line, after introduction of a more negative charge -- blue line.}
\label{fig:Ladowanie}
\end{figure}

The three crystalized forms of n-docosane differ in morphology. PALS experiments indicate that in the case of the powder the electrons are scarcely trapped, whereas dendrites and plates cumulate relatively large amounts of trapped electrons. The VIS absorption spectra of the samples differ from each other and give peaks of absorption which we relate to relatively weakly bound electrons existing in the samples- see fig.~\ref{fig:UVVis}.
\\
%
We were unable to perform structural crystallographic experiments successfully to obtain information on the differences in the crystal structure of our samples. However, some light on the structure was shed after scanning the surface of the samples with the optical confocal microscope and AFM (\emph{Atomic Force Microscope}). The picture of these three types of samples in the scale of tenth of micrometres is shown in fig.~\ref{fig:OpticalPics}.
It seems all the samples have a crystal form (molecule chains are expected to form a molecular crystal where rows of molecular chains are separated by interlamellar gaps with a width of some $\AA$ \cite{Turner}) and some geometrical characteristics of each type are visible. Possibly, the plates and the powder are alike and the differences are the matter of the size of the crystal domains. Both the plates and the powder have a layered structure, where layers spread along the visible plane. The dendrites are considerably different; their crystals are elongated toward one axis whereas their intersections are of hexagonal symmetry. The layered structure of the plates corresponds to structures presented in the literature as stacks of molecules separated by interlamellar gaps. In the case of dendrites, the interlamellar planes lie probably along the sections of the crystal needles.\\
To get insight into the scale of nanometres (the length of the n-docosane chain is about 3~nm), we performed AFM scans for all samples in the scale of 10~$\mu$m (fig.~\ref{fig:AFM10um}) and 1~$\mu$m (fig.~\ref{fig:AFM1um}). At the greatest magnification, it was impossible for us to find any considerable differences between the structures. Although the structure of the powder resembles the plates (fig.~\ref{fig:OpticalPics}), the layered structure is not observed in a more subtle scale (fig.~\ref{fig:AFM10um}). It can also be concluded from fig.~\ref{fig:AFM10um} that a greater partition into crystal domains in the case of the powder is probable (a less ordered structure is recorded here).\\
It is expected that the electrons can be trapped in free volumes, which can be formed by disordered spaces. It can be concluded from figure~\ref{fig:AFM10um} that larger disorder is visible in the scale of tenths of micrometres in the case of the powder. Thus, one could expect that the powder is preferable to contain trapped electrons. The opposite is suggested by PALS. A solution to this contradiction is the fact that the disorder mentioned occurs only in a scale where free volumes formed in this scale are not traps. The differences in the trapping properties of samples could originate from more subtle features e.g. conformation differences or differences in the crystal lattice type (the morphology of the dendrites and plates shown in fig.~\ref{fig:OpticalPics} suggests differences in the crystal structure).
\\
We also recorded the pictures of the samples in polarized light. The incident light was linearly polarized. The second polarizer was set perpendicularly in order to analyse the light transferred from the sample. The result is shown in fig.~\ref{fig:InPolarizedLight} for the three structures. All samples are optically active -- the polarized light coming into the sample changes its polarization. This effect is easy to explain because the chain structure promotes preferences to dipole oscillations along the chain (and thus the absorption of the radiation). This effect will be considered more carefully in the next paragraphs.\\
Furthermore, in the case of the plates, an interference effect between the layers is observed. Here, the colourful patchwork is a result of interference of light with wavelengths suitable for constructive and destructive interference inside the interlayer gaps.
One of the causes of the existence of the VIS peaks mentioned above can be explained taking into account this effect. In the case of the plates, the interference forms regions with partially transmitted light dependent on the wavelength. Transmitted light contains only some wavelengths and due to this, some peaks of absorption could arise. This explanation of the absorption spectrum however does not apply to the dendrites, which do not form layered structures and do not have interference patterns but do have a non-trivial absorption spectrum.\\
Additionally, if the spectra are caused by interference, it is not easy to explain their influence on the magnetic and electric field, which was checked. The results for the dendrites are shown in fig.~\ref{fig:VISchanges}a,b. Here, either the position of peaks or their amplitudes change. The changes are irreversible, i.e. when the field is switched off, the spectrum does not return to the initial shape.
\\
The origin of the presented VIS spectra should be considered taking into account some known physical processes which affect absorption of light in alkanes. The general review of alkane electronic spectra was provided in \cite{Lombos}. Furthermore, the following processes are worth to be mentioned: excitonic transitions \cite{Partridge1, Partridge2, Raymonda67, Kasha59, Morisawa12}, photoconductance \cite{Rosenberg58}, solvatation of electrons \cite{Mozumder,Kroh,Liu,Baxendale}, conformation changes \cite{Maroncelli, Abe, Smith, Morini10}, and the influence of impurities and disorder \cite{Rebane}. All these processes however seem to involve energy originating not from the indicated region and therefore they do not yield the absorption spectrum that we present. The Authors' hypothesis is that electron trapping plays a crucial role here.
The energy spectrum of traps measured previously \cite{Pietrow} suggests that it is possible at room temperature that some amount of trapped electrons have the energy in the scale that we are considering (visible light).\\
An interesting feature of optical spectra is their dependence on polarization of incident light from the spectrometer. The polarization change does influence the spectra considerably. Also here, the place of peaks and their position change -- fig.~\ref{fig:VISchanges}c. The explanation of this fact can be as follows. The optical pictures of our sample presented above suggest that it is preferable that the chains of molecules lie along the glass holder used in the spectrometer. Thus, when changing the position of the polarizer, the polarization orientation can be changed in relation to the orientation of the interlamellar gaps -- fig.~\ref{fig:Polar}. If an electron is assumed to exist (represented by the wave function) in such a free space, it can be excited (removed from this site) by light only if the polarization of the light has a parallel component to the quantization axis (along the width of the gap). In contrast, if the polarization orientation is perpendicular to the gap walls, the photon is unable to excite the electron and transmission of the light is relatively high \cite{Polaryzacja}.
\\
The next argument against the relation of the spectra with interference effects are provided by the ESR (\emph{Electron Spin Resonance}) experiments. They indicate the presence of free radicals in all our samples. The signal is not identical for the samples (fig.~\ref{fig:ESR}). The presence of this signal means that unpaired spins do exist. The spectra are sensitive to the change in the orientation of the sample in the cavity and do not change if the sample is illuminated by light (red, green, blue laser diode, 10 mW was used). The shape of the ESR signal is smeared, which means that the spectrum of energies of free radicals is continuous. The values of the spectroscopic \emph{g} factor for these radicals are from the region 2.00--2.20. We do not know the origin of the radicals formed in the docosane samples but we suggest that they are electrons trapped by the sites of molecules with an induced dipole moment.\\
Factor $g$ for free electrons is $g_e$=2.0023. Organic radicals have the $g$ values in the range 1.99-2.01; for example the $g$ value of the $\cdot$CH$_3$ radical is 2.0026, whereas for DPPH it is 2.0036. Taking into account the $g$ factor values of the docosane samples and free electron as well as organic radicals, we can conclude that radicals in docosane have less ability to move than the other ones in the example.\\
We have
\begin{equation}
h\nu=(g_e+\Delta g)\mu_B B=g\mu_B B,
\end{equation}
where $\Delta g$ includes the effects of local fields and can be positive or negative. It also contains information about the interaction between an electron and an electronic structure of the environment, in particular its local anisotropy. In the case of the local anisotropy of the $g$ factor, the ESR spectrum consists of three peaks at frequencies $g_{xx}B$, $g_{yy}B$ and $g_{zz}B$. Three characteristic peaks are often observed in powders \cite{Dicus10}. The spectra in our case -- fig.~\ref{fig:ESR}, especially for the powder, seem to have the same feature. The anisotropy suggests that the electron interaction depends on the geometry of the site, which can be related to the crystal structures present in a given form of docosane.\\
We made a comparison of the typical ESR spectrum from the dendrites to that for the 'weak pitch' marker. This comparison allows approximating the number of unpaired spins in the sample. Taking into account the number of docosane molecules in this sample, we calculated that there was one unpaired spin per 10$^5$ molecules of docosane.
\\
The resolved hyperfine interactions from abundant protons are possibly not visible in the docosane spectra because of smearing of the hyperfine lines.
\\
Very interesting differences were found in the Raman spectra for our samples. Fig.~\ref{fig:Raman} shows the spectra for the plates (A) and dendrites (B) for several wavelengths of exciting light. The green lines are related to 'resonant' absorption- for the wavelength 514~nm which is highly absorbed (see Vis spectrum, fig.~\ref{fig:UVVis}). The waves 785~nm and 325~nm are weakly absorbed. The Raman scattering depends on the wavelength for both samples. The peak at 2470~cm$^{-1}$ appears only when the sample is excited by wavelength 785~nm. We are not able to relate this peak to a vibration mode of any chemical group.\\
The intensities of almost all peaks are different for the samples. The CH$_2$ vibrations (2900~cm$^{-1}$) are of much higher amplitude in the case of the plates, whereas C-C-C bending vibrations (1000-1500~cm$^{-1}$) are more visible in the dendrites. Similarly, the collective lattice vibrations (100~cm$^{-1}$) are much more preferred in the dendrites. Some shift of peaks is observed in this region (the peak 111~cm$^{-1}$ moves to 109~cm$^{-1}$ in the case of the dendrites).\\
We were able to measure Raman spectra only at room temperature. It is known that at room temperature changes in conformation are highly probable (non-planar conformers are present). It seems that C-C and C-C-C bending indicated by Raman spectroscopy is adequate to these changes. Because Raman oscillations are produced along the molecules -- the preference thereof means lack of non-planar conformers that reduce the sites with elongated dipole moments. Here, these modes are present mostly in the dendrites. Hence, the plates possibly have more non-planar conformers.\\
The most interesting difference is the presence of a wide band with a peak value at ca. 2100~cm$^{-1}$ when the plates are excited by 415~nm. The band could be related to fluorescence, but if so, it should be present in the case of the dendrites (which have the same chemical structure). Additionally, we did not observe fluorescence or phosphorescence in the spectroscopic measurements. The scale of the differences in this part of the spectra is surprising since the samples differ only in a supramolecular scale where the interaction energy is relatively weak. Their weakness is confirmed by no differences in the positions of peaks related to known group vibrations (e.g.~1470, 2900~cm$^{-1}$, \cite{RamanHandbook}). Here, only the amplitudes are different, which is interpreted that the group can be blocked against vibrations but the energy of vibration does not change. The band considered is absent in the case of the dendrites at all. Our hypothesis is that the band observed for the plates is related to weakly bound electrons. This hypothesis is in accordance with the previously derived preference toward formation of non-planar conformers, which form extra free volumes (electron trapping sites). The energy of trapped electrons depends on the site where they adhere. As we have shown in 'Theoretical considerations', the energy may differ and a discrete spectrum caused by transitions would probably not be an adequate description of this type of interactions. Electron vibrations in the traps are less definite than the vibration modes of an electron in a separated chemical bond. That is why bands but not peaks are visible here. The green light seems to be resonant to electron vibrations in the traps.
\\
To check if there are other ways of electron deexcitation than fluorescence and phosphorescence, we carried out photoacoustic measurements of our samples. They showed that only a small amount of absorbed energy was released as heat. Nevertheless, the excitation energy is most likely deposited in lattice vibrations in the case of the powder- fig.~\ref{fig:PAS1}. In the case of the dendrites and plates, this way of deexcitation is less preferable. According to our knowledge, in the case of the powder, the trapped electron density is the lowest (PALS results). Based on our model of electron transfer between traps, it can be claimed that electron transfer from one trap to another after light absorption is less probable (greater value of transition work). Thus, the energy of the excitation is likely to be transformed into the energy of lattice vibrations.
%
%
\\
Finally, to check the evidence for the relation of the Vis spectra to the electrons, we made an attempt to check the sensitivity of the spectra when an extra charge was introduce into the sample (for the sake of simplification of the experiment we switched to heptadecane, C$_{17}$H$_{36}$; the melting point of this hydrocarbon is 20-22$^{\circ}$C, which allows us to deal easily with this hydrocarbon throughout the experiment at ambient temperature). Applying the procedure of charging described in the 'Experimental' we observed an increased absorption around 500-700~nm when the sample was negatively charged (dashed curve in fig.~\ref{fig:Ladowanie}a). If more charge was then introduced, the absorption raised in general and some peaks shifted their position -- fig.~\ref{fig:Ladowanie}b). Furthermore, when the positive charge was introduced to the alkane, no additional absorbance was observed. Even slightly lower absorbance than that in the spectrum for the 'non-charged' sample was observed (shown in fig.~\ref{fig:Ladowanie}a as a thin line). The latter feature could be related to the sucking out some of the quasi-free electrons from the liquid by the attracting force from positive charges.
\\
The peak at about 760~nm is present for all alkanes that we have checked in the liquid phase. It almost does not depend on temperature. The Authors cannot explain its origin.\\
Although the sample was liquid, it is possible to restore electrons in this structure \cite{Kedzia}. The traps are different from those in the solid state. Their origin is probably the solvatation of the electrons by the induced dipoles of the liquid. In the case of a liquid sample, it is probably less easy to keep the electrons inside the sample because solvated electrons can migrate throughout the sample due to the repealing forces.\\
Alkanes are insulators. Therefore, it is probably hardly possible to inject considerably more electrons from the wall to the sample by the method that we have used (charging the external foil of a beaker with the negative charge -- the final part of the procedure of charging described in the 'Experimental'). The amount of free electrons existing in the sample in this case is still small. By the use of parameters of an electroscope that we used to deal with charges, we easily calculated that the amount of electrons we introduced onto the wall was about 10$^9$ whereas the amount of the molecules we had in the sample was 10$^{22}$. Since only a small fraction of electrons was suspended inside the liquid, it seems that the procedure of charging needs to be improved in the future experiments. Nevertheless, it is sufficient to show clearly that the Vis spectra analysed have something in common with extra electrons inside the samples.
\section{Conclusions}
We examined the UV-Vis spectra obtained from the point of view of recent theories concerning the electronic spectra in hydrocarbons. We have shown that the spectra are sensitive to charging the sample and we have demonstrated with the use of other techniques that alkane crystals accumulate some amount of quasi-free electrons. Their amount as well as bound energy depends on the crystal morphology. We related the peaks in the VIS spectra to these electrons and explained their dependence on polarization of light exciting them. This explanation supports the idea that these electrons are located in free volumes following the symmetry of the crystal, i.e. interlamellar volumes and kinks. We have shown that Raman spectra indicate preferences to formation of conformers (and thus free volumes) in plates more than in the other structures. The photospectroscopy of these electrons can be a probe of structural characteristics of samples.
\section{Acknowledgements}
The authors want to thank Dr. Kasia Mill (University of Life Sciences in Lublin) for taking the pictures of the crystal in polarized light. We thank Prof.~S.~Krawczyk and Prof.~W.I.~Gruszecki (Physics Dept. MCS University, Lublin) for sharing some equipment from their labs and Prof.~J.~K\k{e}dzia (Opole University of Technology) for literature recommendation.
\bibliographystyle{unsrt}
\bibliography{references}
\end{document}